\documentclass[aps,superscriptaddress,prc,twocolumn,nofootinbib]{revtex4}

\usepackage{graphicx}
\usepackage{epstopdf}
\usepackage{subfigure}
\usepackage{epsfig}
\usepackage{amsmath,amssymb,amsfonts}
\usepackage{color}

\begin{document}

\title{Energy loss, hadronization and hadronic interactions of heavy flavors in relativistic heavy-ion collisions}

\author{Shanshan Cao}
\affiliation{Nuclear Science Division, Lawrence Berkeley National Laboratory, Berkeley, CA 94720, USA}
\affiliation{Department of Physics, Duke University, Durham, NC 27708, USA}

\author{Guang-You Qin}
\affiliation{Institute of Particle Physics and Key Laboratory of Quark and Lepton Physics (MOE), Central China Normal University, Wuhan, 430079, China}

\author{Steffen A. Bass}
\affiliation{Department of Physics, Duke University, Durham, NC 27708, USA}

\date{\today}


\begin{abstract}

We construct a theoretical framework to describe the evolution of heavy flavors produced in relativistic heavy-ion collisions. The in-medium energy loss of heavy quarks is described using our modified Langevin equation that incorporates both quasi-elastic scatterings and the medium-induced gluon radiation. The space-time profiles of the fireball are described by a (2+1)-dimensional hydrodynamics simulation. A hybrid model of fragmentation and coalescence is utilized for heavy quark hadronization, after which the produced heavy mesons together with the soft hadrons produced from the bulk QGP are fed into the hadron cascade UrQMD model to simulate the subsequent hadronic interactions. We find that the medium-induced gluon radiation contributes significantly to heavy quark energy loss at high $p_\mathrm{T}$; heavy-light quark coalescence enhances heavy meson production at intermediate $p_\mathrm{T}$; and scatterings inside the hadron gas further suppress the $D$ meson $R_\mathrm{AA}$ at large $p_\mathrm{T}$ and enhance its $v_2$. Our calculations provide good descriptions of heavy meson suppression and elliptic flow observed at both the LHC and RHIC.

\end{abstract}

\maketitle


\section{Introduction}
\label{sec:introduction}

The primary purpose of performing relativistic heavy-ion collisions at the Large Hadron Collider (LHC) and the Relativistic Heavy-Ion Collider (RHIC) is to study the properties of QCD matter under extreme conditions such as high temperatures and densities. It has now been well established that a new state of matter, known as the strongly interacting quark-gluon plasma (sQGP) \cite{Gyulassy:2004zy,Shuryak:2004cy}, is created in these energetic nuclear collisions. This highly excited state of matter is composed of color de-confined quarks and gluons, and displays properties of a nearly perfect fluid such as the strong collective flow observed for the produced hadrons \cite{Adare:2011tg, Adamczyk:2013waa, ATLAS:2012at}. Relativistic hydrodynamic models successfully describe the space-time evolution of the strongly coupled QGP fireballs \cite{Teaney:2000cw,Huovinen:2001cy,Nonaka:2006yn,Song:2007fn,Luzum:2008cw,Qiu:2011hf,Pang:2012uw, Gale:2012rq}, from which it is found that the value of the shear viscosity to entropy density ratio $\eta/s$ of the produced QGP is small.

Apart from studying soft hadrons emitted from the QGP directly, an alternative way to study the transport properties of the QGP is through the investigation of the modification to the properties of energetic partons that travel through the produced hot and dense medium.  One of the promising candidates is heavy quarks. Owing to their large masses, the thermal production of heavy quarks from the QGP fireball is significantly suppressed, thus the majority of them are produced during the primordial stage of the collision via hard scatterings. These heavy quarks then propagate through the medium, and can probe the whole evolution history of the QGP. Over the past decade, experimental observations at both the LHC and RHIC have revealed many interesting and sometimes surprising observations of heavy flavor hadrons and their decay electrons, such as the small values of their nuclear modification factors $R_\mathrm{AA}$ and the large values of their elliptic flow coefficients $v_2$ which are almost comparable to those of light hadrons \cite{Adare:2010de,Adare:2014rly,Tlusty:2012ix,Adamczyk:2014uip,Grelli:2012yv,Abelev:2013lca}. This seems contradictory to the earlier expectation from the mass hierarchy of parton energy loss and still remains a challenge for us to fully understand.

Various transport models have been constructed to study the heavy quark motion inside dense nuclear matter, such as the parton cascade model based on the Boltzmann equation \cite{Molnar:2006ci,Zhang:2005ni,Uphoff:2011ad,Uphoff:2012gb} and the linearized Boltzmann approach coupled to a hydrodynamic background \cite{Gossiaux:2010yx,Nahrgang:2013saa}. In the limit of small momentum transfer, the multiple scatterings of heavy quarks inside a thermalized medium can be treated as Brownian motion, and the Boltzmann equation for quasi-elastic scatterings is then reduced to the Fokker-Plank equation which can then be stochastically realized by the Langevin equation. Many Langevin-based transport models \cite{Svetitsky:1987gq,Moore:2004tg, Akamatsu:2008ge,He:2011qa,Young:2011ug,Alberico:2011zy,Lang:2012cx,Cao:2011et,Cao:2012jt} have been developed to study the collisional energy loss of heavy quarks and have been shown to be successful in describing experimental data in the low transverse momentum $p_\mathrm{T}$ region where the phase space for the medium-induced gluon radiation is restricted by the large masses of heavy quarks, i.e., the ``dead cone effect" \cite{Dokshitzer:2001zm,Abir:2012pu}. However, LHC experiments now enable us to observe heavy meson spectra up to 30~GeV. At such high $p_\mathrm{T}$, even heavy quarks become ultra-relativistic and therefore the radiative energy loss should no longer be neglected. In our previous work \cite{Cao:2012au,Cao:2013ita}, the classical Langevin equation is modified such that quasi-elastic scattering and medium-induced gluon radiation can be incorporated simultaneously. In this study we will continue utilizing this improved Langevin approach for the in-medium evolution of heavy quarks.

A dedicated description of the heavy quark energy loss inside the QGP is crucial for solving the ``heavy flavor puzzle", but has yet to be accomplished. It has been pointed out that details in the hadronization process may have a strong impact on the observed heavy meson spectra \cite{Lin:2003jy,Greco:2003vf,Oh:2009zj,vanHees:2005wb,vanHees:2007me,He:2011qa}. The influence of hadronic interactions after the QGP decays on heavy meson observables has also been explored in Refs. \cite{He:2011yi,He:2014cla} and has been shown to be non-negligible. In this work, we will develop a hybrid model of fragmentation plus coalescence to describe the heavy quark hadronization process. The momentum dependence of the relative probability between fragmentation and coalescence will be calculated according to the Wigner functions in an instantaneous coalescence model. This coalescence model was first proposed for the production of light hadrons out of QGP fireballs \cite{Dover:1991zn,Fries:2003kq,Greco:2003mm,Chen:2006vc}, and then applied to the production of heavy flavor hadrons in nuclear collisions \cite{Lin:2003jy,Greco:2003vf,Oh:2009zj} and recently to partonic jet hadronization \cite{Han:2012hp} as well. This model does not require the thermalization of the recombining partons and is easily extensible to simultaneously include various meson and baryon species, allowing for the normalization of the total coalescence probability over all possible hadronization channels. Based on our previous study \cite{Cao:2013ita}, we will further develop this hybrid hadronization model such that it is applicable to arbitrary local flow velocities of the QGP background. In addition, the rescattering of heavy mesons inside a hadron gas after hadronization will also be incorporated in this work by utilizing the Ultra-relativistic Quantum Molecular Dynamic Model (UrQMD) \cite{Bass:1998ca}, and the effect of hadronic interactions on the observed heavy meson spectra will be investigated in detail.

The paper is organized as follows. In Sec. \ref{sec:HQenergyloss}, we will present how the classical Langevin equation is modified to simultaneously incorporate collisional and radiative energy loss of heavy quarks and how the simulation of the heavy quark evolution in a dynamic QGP medium is implemented. In Sec. \ref{sec:HFhadr}, we develop a hybrid model of fragmentation and coalescence to describe the hadronization of heavy quarks. With that hadronization model, we present numerical results of heavy meson suppression and anisotropic flow and compare to experimental data at the LHC and RHIC. In Sec. \ref{sec:hadronicInteraction}, we will discuss how the hadronic rescattering of heavy mesons is simulated within the UrQMD model and its effect on the observed heavy meson $R_\mathrm{AA}$ and $v_2$. We will summarize and discuss future developments in Sec. \ref{sec:summary}.

\section{Heavy quark energy loss in QGP matter}
\label{sec:HQenergyloss}

\subsection{A modified Langevin equation}
\label{subsec:Langevin}

During their propagation through a thermalized QCD matter, heavy quarks lose energy via both quasi-elastic scatterings with light patrons in the medium and gluon radiation induced by multiple scatterings. In this work, we utilize the following modified Langevin equation \cite{Cao:2013ita} that simultaneously incorporates these two processes to describe the time evolution of energy and momentum of heavy quarks while they traverse the QGP matter:
\begin{equation}
\label{eq:modifiedLangevin}
\frac{d\vec{p}}{dt}=-\eta_D(p)\vec{p}+\vec{\xi}+\vec{f}_g.
\end{equation}
In Eq. (\ref{eq:modifiedLangevin}), the first two terms on the right-hand side are inherited from the classical Langevin equation and represent the drag force and the thermal random force experienced by a heavy quark while it diffuses inside a thermal medium due to multiple scatterings. For a minimal model, we assume the thermal force $\vec{\xi}$ is independent of the heavy quark momentum and satisfies the correlation relation of a white noise $\langle\xi^i(t)\xi^j(t')\rangle=\kappa\delta^{ij}\delta(t-t')$, in which $\kappa$ denotes the momentum diffusion coefficient of heavy quarks and is related to the spatial diffusion coefficient $D$ via $D\equiv T/[M\eta_D(0)]=2T^2/\kappa$ if the fluctuation-dissipation theorem $\eta_D(p)=\kappa/(2TE)$ is respected.

Apart from the above two forces resulting from quasi-elastic scatterings, an additional term $\vec{f}_g=-d\vec{p}_g/dt$ is introduced into Eq. (\ref{eq:modifiedLangevin}) to describe the recoil force exerted on heavy quarks while experiencing medium-induced gluon radiation, where $\vec{p}_g$ is the momentum of the radiated gluon. The probability of gluon radiation during the time interval $[t,t+\Delta t]$ is determined based on the average number of radiated gluons in this $\Delta t$:
\begin{equation}
\label{eq:gluonnumber}
P_\mathrm{rad}(t,\Delta t)=\langle N_\mathrm{g}(t,\Delta t)\rangle = \Delta t \int dxdk_\perp^2 \frac{dN_\mathrm{g}}{dx dk_\perp^2 dt}.
\end{equation}
As long as $\Delta t$ is chosen sufficiently small, $\langle N_\mathrm{g}(t,\Delta t)\rangle$ is less than 1 and can be interpreted as a probability. In this study, the gluon distribution function in Eq. (\ref{eq:gluonnumber}) is adopted from the higher-twist calculation for the medium-induced gluon radiation -- the distribution function of gluons radiated from a massless parton is calculated in Refs. \cite{Guo:2000nz,Majumder:2009ge} and its modification due to the mass effect of a heavy quark is introduced by Ref. \cite{Zhang:2003wk}:
\begin{eqnarray}
\label{eq:gluondistribution}
\frac{dN_\mathrm{g}}{dx dk_\perp^2 dt}=\frac{2\alpha_s  P(x)\hat{q} }{\pi k_\perp^4} {\sin}^2\left(\frac{t-t_i}{2\tau_f}\right)\left(\frac{k_\perp^2}{k_\perp^2+x^2 M^2}\right)^4,
\end{eqnarray}
in which $x$ is the fractional energy taken by the emitted gluon from the heavy quark, and $k_\perp$ is the  transverse momentum of the gluon. $\alpha_s$ is the strong coupling constant, $P(x)$ is the gluon splitting function and $\tau_f$ is the formation time of the gluon defined as $\tau_f={2Ex(1-x)}/{(k_\perp^2+x^2M^2)}$ with $E$ and $M$ being the energy and mass of heavy quarks. Note that the multiplicative term at the end of Eq. (\ref{eq:gluondistribution}) is known as the ``dead cone factor", signifying the mass dependence of the radiative energy loss of hard parton. In Eq. (\ref{eq:gluondistribution}), $\hat{q}$ is the gluon transport coefficient and may be related to the above mentioned quark diffusion coefficient $\kappa$ via $\hat{q}=2\kappa C_A/C_F$. Therefore, in our calculations there is only one free parameter in the modified Langevin equation [Eq. (\ref{eq:modifiedLangevin})]. To obtain the best description of heavy flavor observables at the LHC and the RHIC, as will be shown in Sec. \ref{sec:HFhadr} and Sec. \ref{sec:hadronicInteraction}, the spatial diffusion coefficient of heavy quark $D(2\pi T)$ is chosen around $5\sim6$, which is equivalent to $\hat{q}/T^3$ around $9.4\sim11.3$ for gluons (or $4.2\sim5.0$ for quarks), consistent with the value extracted by the JET Collaboration via fitting the experimental data using various jet energy loss models \cite{Burke:2013yra}.

When simulating the radiative energy loss of heavy quarks, a lower cut-off of radiated gluon energy $\omega_0=\pi T$ is imposed to take into account the balance between gluon emission and absorption processes. Below $\omega_0$, the gluon radiation is disabled and the evolution of heavy quarks with low energies is entirely controlled by quasi-elastic scatterings. For this reason, $x\in [\pi T/E,1]$ is used when calculating the gluon radiation probability in Eq. (\ref{eq:gluonnumber}). With this treatment, we have verified in our previous work \cite{Cao:2013ita} that the thermal equilibration of heavy quarks can be approached after sufficiently long evolution time although the exact fluctuation-dissipation relation may not be guaranteed due to the lack of the gluon absorption process. A more detailed discussion of this approach and possible improvements towards a more rigorous treatment of detailed balance between gluon emission and absorption were discussed Reference. \cite{Cao:2013ita}. And alternative approaches of including the radiative energy loss into the Langevin framework can be found in Refs. \cite{Gossiaux:2006yu,Das:2010tj}.

\subsection{Heavy quark evolution in a realistic medium}
\label{subsec:simulation}

To study the heavy flavor spectra produced in realistic heavy-ion collisions, we couple the above modified Langevin equation to an expanding QGP medium that is simulated with a (2+1)-dimensional viscous hydrodynamic model developed in Refs. \cite{Song:2007fn,Song:2007ux,Qiu:2011hf}. In this work, we utilize the code version and parameter values provided by Ref. \cite{Qiu:2011hf}. The hydrodynamic simulation generates the space-time evolution of the local temperature and flow velocity profiles of the QGP fireball created in relativistic nuclear collisions. For every time step of the Langevin evolution, we first boost each heavy quark into the local rest frame of the fluid cell through which it propagates. In the rest frame of fluid cell, the energy and momentum of a given heavy quark are updated using Eq. (\ref{eq:modifiedLangevin}) before it is boosted back to the global center of mass frame.

The hydrodynamical evolution of the bulk matter is initialized with either the Monte Carlo (MC) Glauber or the Kharzeev-Levin-Nardi (KLN) parametrization of the Color Glass Condensate (CGC) model for its entropy density distribution. To best describe the spectra of soft hadrons emitted from the QGP fireballs, for both the RHIC and the LHC environments, the starting time of the QGP evolution has been set as $\tau_0=0.6$~fm/$c$ and the shear-viscosity-to-entropy-density ratio ($\eta/s$) has been tuned as 0.08 when the Glauber initial condition is used and 0.20 when KLN is used. In this work, a smooth initial condition is utilized for the bulk matter. Possible effects of the the initial state fluctuation on heavy flavor observables have been discussed in our earlier study \cite{Cao:2014fna}. For heavy quarks, we use the MC Glauber model to initialize their production positions and the leading-order perturbative QCD (pQCD) approach \cite{Combridge:1978kx} to calculate their initial momentum space distribution. We have included the pair production process ($gg\rightarrow Q\bar{Q}$ and $q\bar{q}\rightarrow Q\bar{Q}$) and the flavor excitation process ($gQ\rightarrow gQ$ and $g\bar{Q}\rightarrow g\bar{Q}$) in calculating the initial $p_\mathrm{T}$ spectra of heavy quarks. 
The gluon splitting process ($g\rightarrow Q\bar{Q}$) has been recently discussed in Ref. \cite{Kang:2011rt} and will be investigated in a follow-up study. 
These pQCD calculations are at the partonic level. To calculate the cross sections of heavy quark production in nuclear collisions, we adopt CTEQ for the parton distribution functions \cite{Lai:1999wy} and include the nuclear shadowing/anti-shadowing effect in heavy-ion collisions using the EPS09 parametrization \cite{Eskola:2009uj}.

\begin{figure}[tb]
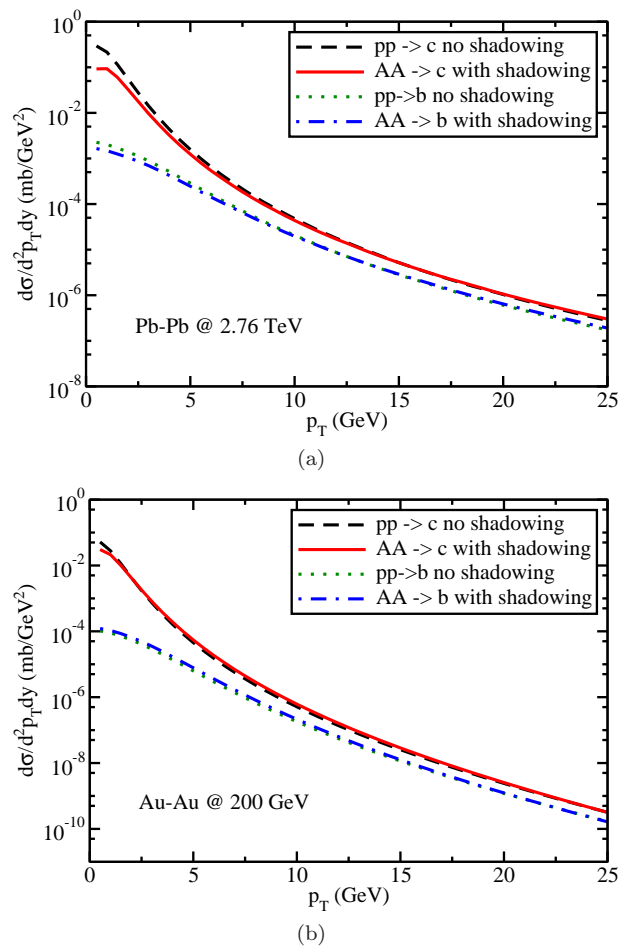

 \subfigure[]{\label{fig:LHCppvsAA}
      \epsfig{file=plot-LHCppvsAA.eps, width=0.45\textwidth, clip=}}
 \subfigure[]{\label{fig:RHICppvsAA}
      \epsfig{file=plot-RHICppvsAA.eps, width=0.45\textwidth, clip=}}
 \caption{(Color online) The initial heavy flavor spectra from the leading-order pQCD calculation with and without the nuclear shadowing effect (EPS09), (a) for the LHC and (b) for the RHIC experiments.}
 \label{fig:initialSpectra}
 \end{figure}

In Fig. \ref{fig:initialSpectra}, we show the $p_\mathrm{T}$ spectra of initial heavy quarks at both LHC and RHIC energies, for proton-proton collisions and binary collision number scaled nucleus-nucleus collisions. The influence of the nuclear shadowing/anti-shadowing effect in the initial state on heavy quark spectra can be clearly observed in the figures: it reduces the production rate of charm quarks at low $p_\mathrm{T}$ but slight enhances it at high $p_\mathrm{T}$; the effect is stronger at the LHC energy than at the RHIC. 
For bottom quarks, the production of the low $p_\mathrm{T}$ bottom quarks is decreased at the LHC energy but slightly enhanced at the RHIC when initial state effects are included. 
Such effects will have significant impact on the nuclear modification factor $R_\mathrm{AA}$ of heavy mesons observed in the final state as will be shown later in Sec. \ref{sec:HFhadr}. 
The calculated spectra are used to sample the initial $p_\mathrm{T}$ distributions of heavy quarks. 
Their initial rapidity distributions are taken to be uniform around the central region ($-1<\eta<1$).

In simulating the evolution of heavy quarks, they are assumed to stream freely first from their production vertices in hard collisions to $\tau_0=0.6$~fm/$c$, the initial time at which the hydrodynamical evolution commences. The possible energy loss in the pre-equilibrium stage has been neglected which is expected to give only a small contribution to the final state spectra, given its short period of time compared to the much longer evolution of the QGP fireball.

 \begin{figure}[tb]
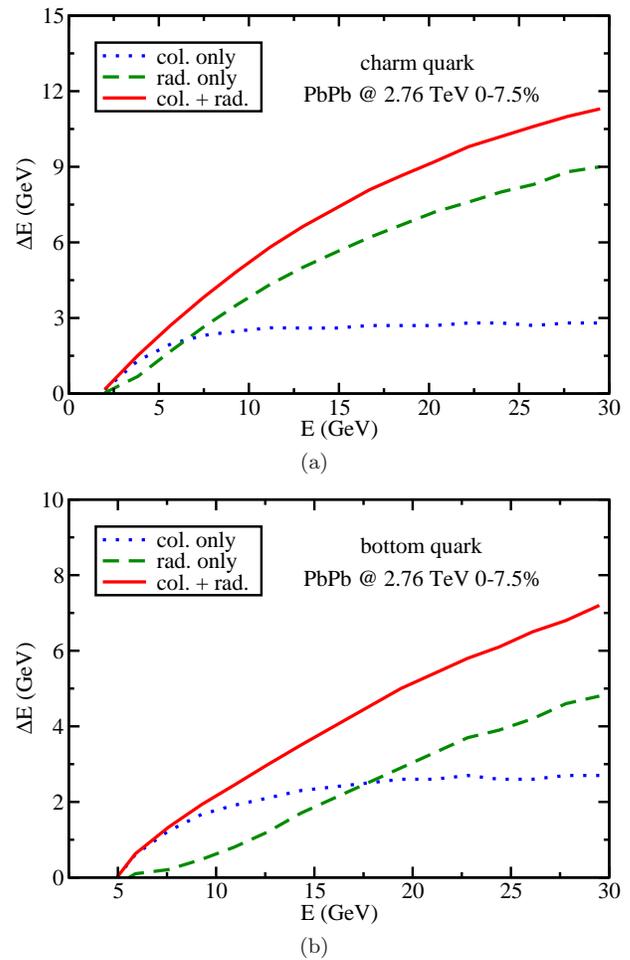

 \subfigure[]{\label{fig:eloss_cD5-LHC-0-7d5}
      \epsfig{file=plot-eloss_cD5-LHC-0-7d5.eps, width=0.45\textwidth, clip=}}
 \subfigure[]{\label{fig:eloss_bD5-LHC-0-7d5}
      \epsfig{file=plot-eloss_bD5-LHC-0-7d5.eps, width=0.45\textwidth, clip=}}
 \caption{(Color online) Comparison of energy loss between different mechanisms: (a) for charm quark and (b) for bottom quark.}
 \label{fig:energyLoss}
 \end{figure}

With the above setup, we can investigate how heavy quarks evolve inside QGP matter and lose their energy. In Fig. \ref{fig:energyLoss}, we calculate the average energy loss of charm and bottom quarks as a function of their initial energy after they traverse a realistic QGP medium created by central Pb-Pb collisions at the LHC energy. The contributions from different energy loss mechanisms are compared. As shown by the figures, for both charm and bottom quarks, quasi-elastic scatterings dominate their energy loss while their initial energies are small, however, medium-induced gluon radiation dominates in the high energy regimes. The crossing point is around 7~GeV for charm quarks, and increases to 18~GeV for bottom quarks due to the greater suppression of gluon radiation by its larger masses. These results indicate that collisional energy loss alone may provide reasonable descriptions for the heavy flavor observables in the low $p_\mathrm{T}$ region as those measured at RHIC, but will become insufficient when we extend to higher $p_\mathrm{T}$ such as those reached by the LHC experiments.

\section{Heavy flavor hadronization}
\label{sec:HFhadr}

In the previous section, we studied the initial production of heavy quarks in heavy-ion collisions and their energy loss inside a QGP medium. Around the critical temperature $T_\mathrm{c}=165$~MeV, both the bulk matter of the QGP fireball and heavy quark should hadronize into color neutral bound states. For the bulk matter, we utilize the numerical tool ``iSS" \cite{Shen:2014vra} based on the Cooper-Frye formula \cite{Cooper:1974mv} to obtain soft hadrons from the hydrodynamic medium. For heavy quarks, we follow our previous work \cite{Cao:2013ita} and develop a hybrid model of fragmentation and coalescence to describe their hadronization. After heavy mesons are obtained from heavy quarks, we may directly compare their suppression and collective flow coefficients with experimental data from both the LHC and RHIC.

\subsection{A hybrid model of fragmentation and coalescence}
\label{subsec:hybrid}

Between the two typical in-medium hadronization processes, fragmentation and heavy-light quark coalescence, of heavy quarks into heavy flavor hadrons, the former dominates the high momentum regimes while the latter becomes important at low momenta. The momentum dependence of the relative probability between these two mechanisms can be determined by the Wigner function in the instantaneous coalescence model \cite{Oh:2009zj}. With the knowledge of this probability, spectra of heavy mesons formed from the heavy-light quark coalescence can be directly calculated within the coalescence model, while those from the fragmentation process can be obtained from the \textsc{Pythia} simulation \cite{Sjostrand:2006za}. In our previous study \cite{Cao:2013ita}, a hybrid model of fragmentation and coalescence was established for the heavy flavor hadronization. In this work, we will further develop this hadronization model so that the effect of the local flow of an expanding medium on hadronization can be conveniently taken into account.

In the instantaneous coalescence model, the momentum spectra of produced mesons and baryons are given as follows,
\begin{eqnarray}
\label{eq:recombMeson}
\frac{dN_M}{d^3p_M} \!\!&=&\!\!\! \int d^3p_1 d^3p_2 \frac{dN_1}{d^3p_1} \frac{dN_2}{d^3p_2}f^W_M(\vec{p}_1,\vec{p}_2)\delta(\vec{p}_M-\vec{p}_1-\vec{p}_2) \nonumber \\
\frac{dN_B}{d^3p_B} \!\!&=&\!\!\! \int d^3p_1 d^3p_2 d^3p_3 \frac{dN_1}{d^3p_1} \frac{dN_2}{d^3p_2} \frac{dN_3}{d^3p_3}f^W_B(\vec{p}_1,\vec{p}_2,\vec{p}_3) \nonumber \\  && \times \delta(\vec{p}_M-\vec{p}_1-\vec{p}_2-\vec{p}_3).
\end{eqnarray}
$dN_i/d^3p_i$ denotes the momentum distribution of the $i$-th valence quark in the produced hadron. The spectra of heavy quarks can be directly obtained after they traverse the QGP fireball within our modified Langevin evolution.  Light quarks are assumed thermal in the local rest frame of the expanding medium:
\begin{equation}
\label{eq:FermiDirac}
\frac{dN_q}{d^3p}=\frac{g_q V}{e^{\sqrt{p^2+m^2}/T}+1},
\end{equation}
in which $g_q=6$ is the statistic factor that takes into account spin and color degeneracy of quark, and for simplicity a uniform distribution in the position space is assumed inside a volume $V$. In Eq. (\ref{eq:recombMeson}), one key ingredient of the coalescence model is the Wigner function $f^W$ which denotes the probability for the two or three quarks to combine. For a two-body system, the Wigner function can be written as
\begin{equation}
\label{eq:WignerMeson}
f_M^W(\vec{r},\vec{q})\equiv g_M \int d^3 r' e^{-i\vec{q}\cdot\vec{r}'}\phi_M(\vec{r}+\frac{\vec{r}'}{2})\phi^*_M(\vec{r}-\frac{\vec{r}'}{2}),
\end{equation}
in which $g_M$ denotes the degrees of freedom (spin and color) of the formed meson and the variables $\vec{r}$ and $\vec{q}$ are the relative position and momentum of the two particles defined in the two-body center-of-mass frame, i.e., the rest frame of the produced meson: 
\begin{eqnarray}
\vec{r} \equiv \vec{r}_1^\mathrm{\,cm}-\vec{r}_2^\mathrm{\,cm},\quad 
\vec{q} \equiv \frac{E_2^\mathrm{\,cm} \vec{p}_1^\mathrm{\,cm}-E_1^\mathrm{\,cm}\vec{p}_2^\mathrm{\,cm}}{E_1^\mathrm{\,cm}+E_2^\mathrm{\,cm}}.
\end{eqnarray} 
Note that the heavy and light quarks are first boosted into their center-of-mass frame in which their coalescence probability is then calculated. In Eq. (\ref{eq:WignerMeson}), $\phi_M$ represents the meson wavefunction, which is approximated by the ground state wavefunction of a simple harmonic oscillator: $\mathrm{exp}[-r^2/(2\sigma^2)]/(\pi \sigma^2)^{3/4}$. Here, the width $\sigma$ is related to the angular frequency of the oscillator $\omega$ via $\sigma\equiv 1/\sqrt{\mu\omega}$, with $\mu\equiv m_1m_2/(m_1+m_2)$ being the reduced mass of the two-body system. With these setups, we may average over the position space of Eq. (\ref{eq:WignerMeson}) and obtain the momentum space Wigner function of the produced meson:
\begin{equation}
\label{eq:momentumWigner}
f^W_M(q^2) = g_M \frac{(2\sqrt{\pi}\sigma)^3}{V} e^{-q^2 \sigma^2}.
\end{equation}
The above procedure can be straightforwardly generalized to a three-body system for baryon formation by combining two quarks first and then combining their center of mass with the third quark:
\begin{equation}
\label{eq:baryonWigner}
f^W_B(q_1^2,q_2^2) = g_B \frac{(2\sqrt{\pi})^6(\sigma_1\sigma_2)^3}{V^2} e^{-q_1^2 \sigma_1^2-q_2^2\sigma_2^2},
\end{equation}
with $\vec{q}_1$ and $\vec{q}_2$ as the relative momenta defined in the rest frame of the produced baryon 
\begin{eqnarray}
&& \vec{q}_1\equiv \frac{E_2^\mathrm{\,cm}\vec{p}_1^\mathrm{\,cm}-E_1^\mathrm{\,cm}\vec{p}_2^\mathrm{\,cm}}{E_1^\mathrm{\,cm}+E_2^\mathrm{\,cm}}, 
\nonumber\\
&& \vec{q}_2\equiv \frac{E_3^\mathrm{\,cm}(\vec{p}_1^\mathrm{\,cm}+\vec{p}_2^\mathrm{\,cm})-(E_1^\mathrm{\,cm}+E_2^\mathrm{\,cm})\vec{p}_3^\mathrm{\,cm}}{E_1^\mathrm{\,cm}+E_2^\mathrm{\,cm}+E_3^\mathrm{\,cm}}, 
\end{eqnarray}
and $\sigma_i=1/\sqrt{\mu_i \omega}$ as the width parameter with $\mu_1\equiv m_1m_2/(m_1+m_2)$ and $\mu_2\equiv (m_1+m_2)m_3/(m_1+m_2+m_3)$. In the calculations, the thermal mass is taken as 300~MeV for $u$ and $d$ quarks and 475~MeV for $s$ quarks. Heavy quarks, on the other hand, are not required to be thermal, and their masses are taken as 1.27~GeV for $c$ and 4.19~GeV for $b$ quarks. Contribution from thermal gluons is also incorporated in this coalescence model: they are split into light quark pairs first and then combine with heavy quarks to form heavy flavor hadrons.

\begin{figure}[tb]
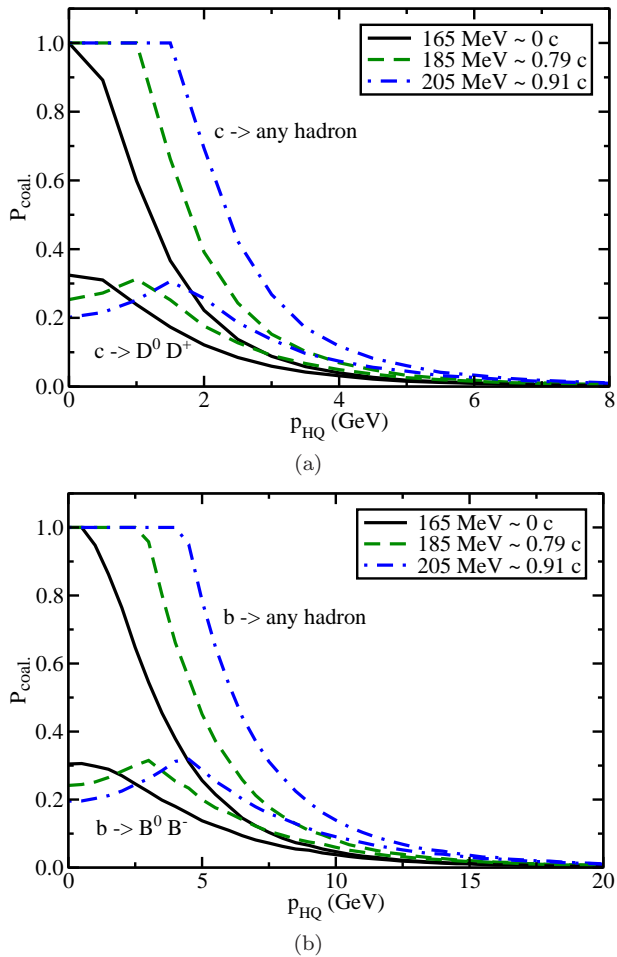

 \subfigure[]{\label{fig:Pcoal_c}
      \epsfig{file=plot-Pcoal_c.eps, width=0.45\textwidth, clip=}}
 \subfigure[]{\label{fig:Pcoal_b}
      \epsfig{file=plot-Pcoal_b.eps, width=0.45\textwidth, clip=}}
 \caption{(Color online) The momentum dependence of the coalescence probabilities at different flow velocities: (a) for charm quark and (b) for bottom quark.}
 \label{fig:Pcoal}
 \end{figure}

We use Eqs.(\ref{eq:momentumWigner}) and (\ref{eq:baryonWigner}) to evaluate the momentum dependence of heavy-light quark coalescence probabilities at the critical temperature $T_\mathrm{c}$, as shown in Fig. \ref{fig:Pcoal}. In principle, the oscillator frequency $\omega$ in these Wigner functions can be calculated from the charge radius of the hadrons and should depend on the hadron species. Here for a minimal model, we adopt an average value 0.215~GeV for all $c$-hadrons and 0.102~GeV for $b$-hadrons. These two parameters are obtained by requiring the coalescence probability through all possible hadronization channels to be unity for a zero momentum heavy quark in a static medium at $T_\mathrm{c}$ since it is not sufficiently energetic to fragment \cite{Oh:2009zj}, as can be seen in Fig. \ref{fig:Pcoal}. In our calculations, all major hadron channels are incorporated, including the ground states and the first excited states of $D$/$B$ mesons, $\Lambda_Q$, $\Sigma_Q$, $\Xi_Q$ and $\Omega_Q$.

After the $\omega$ parameters are evaluated in a static medium according to the above normalization procedure, the Wigner functions are determined. 
For heavy quarks in an expanding medium, we adopt an effective temperature method \cite{Oh:2009zj,Oh:2007vf} to calculate the effective temperature of a fluid cell with a non-zero velocity due to a blue shift effect as follows:
\begin{equation}
\label{eq:effectiveT}
\sum_{q,g}\int d^3p \frac{g_{q,g} V}{e^{E_{q,g}/T_\mathrm{eff}}\pm1}=\sum_{q,g}\int d^3p \frac{g_{q,g} V}{e^{p_{q,g}\cdot u/T_\mathrm{c}}\pm1},
\end{equation}
in which $u$ is the 4-velocity of the fluid cell. This effective temperature $T_\mathrm{eff}$ is then utilized in the thermal distributions of light partons and the coalescence probability inside a moving fluid cell is calculated according to Eqs.(\ref{eq:momentumWigner}) and (\ref{eq:baryonWigner}). If the obtained value of the coalescence probability is greater than unity at low momenta, it is taken as unity.

In Fig. \ref{fig:Pcoal}, we show our calculations of the coalescence probabilities for both charm and bottom quarks as functions of their momenta either through heavy meson channel alone ($D$/$B$ meson), or to any possible hadrons (summing over all hadron channels under consideration). Three different values of the fluid flow velocity, which correspond to three different effective temperatures are compared. One can observe that the coalescence probability generally decreases with the increase of heavy quark momentum, and a larger fluid velocity leads to a higher effective temperature and therefore an enhanced coalescence probability. Furthermore, for the same momentum, bottom quarks have larger probability to coalesce with light quarks than charm quarks do, due to the larger mass (or smaller velocity) of the bottom quarks inside a QGP medium.

In Fig. \ref{fig:Pcoal} we divide the hadronization of heavy quarks into three regimes: coalescence with light quarks to $D$ or $B$ mesons, coalescence to other hadron channels, and fragmentation. After its evolution through the QGP matter, if a charm or bottom quark is selected for coalescence into a $D$ or $B$ meson, a light quark or anti-quark is generated according to thermal distribution at $T_\mathrm{eff}$ in the local rest frame of the fluid cell, and then boosted to the lab frame to combine with the given heavy quark according to the probability governed by Eq. (\ref{eq:momentumWigner}). If they do not combine, another light parton is generated until a meson is formed. On the other hand, if a heavy quark is selected to fragment based on the probability in Fig. \ref{fig:Pcoal}, its fragmentation is implemented via \textsc{Pythia} in which the relative ratios between different hadron channels are properly calculated and normalized.

\begin{figure}[tb]
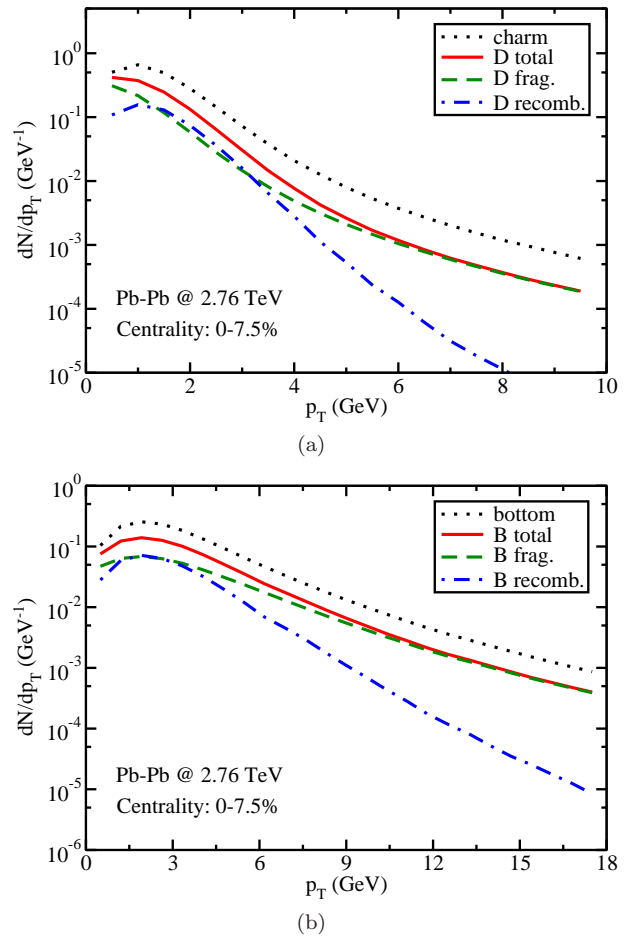

 \subfigure[]{\label{fig:DSpectra}
      \epsfig{file=plot-DSpectra.eps, width=0.45\textwidth, clip=}}
 \subfigure[]{\label{fig:BSpectra}
      \epsfig{file=plot-BSpectra.eps, width=0.45\textwidth, clip=}}
 \caption{(Color online) Comparison between the contributions from different hadronization mechanisms to the (a) $D$ and (b) $B$ meson spectra (normalized to one heavy quark).}
 \label{fig:HMSpectra}
 \end{figure}

Using this hybrid model of hadronization, we may compare the relative contributions from coalescence and fragmentation mechanisms to heavy meson production in relativistic nuclear collisions. As can be observed in Fig. \ref{fig:HMSpectra}, after charm and bottom quarks traverse a realistic QGP medium created in central Pb-Pb collisions at the LHC energy, their hadronization to $D$/$B$ mesons are dominated by fragmentation at high $p_\mathrm{T}$ but is significantly enhanced by heavy-light quark coalescence at intermediate $p_\mathrm{T}$. Since the coalescence mechanism combines a thermal parton and a heavy quark, the spectrum of $D$/$B$ mesons is shifted to the larger momentum regime compared to the original charm/bottom quark distribution. Therefore, its contribution to the production of heavy mesons at low $p_\mathrm{T}$ is not as significant as that at intermediate $p_\mathrm{T}$. Furthermore, as already seen in Fig. \ref{fig:Pcoal}, due to the larger masses and thus smaller velocities of $b$ quarks than $c$ quarks, the coalescence mechanism dominates a wider $p_\mathrm{T}$ range for $B$ meson production than for $D$ meson production.

\subsection{Heavy flavor suppression and collective flow}
\label{subsec:RaaV2}

With our modified Langevin equation for the in-medium evolution of open heavy quark and the above hybrid model of fragmentation and coalescence for heavy quark hadronization, we are able to calculate the suppression and elliptic flow coefficients of heavy flavor hadrons and compare them with experimental data from the LHC and RHIC. Discussions on the additional variation of the heavy flavor observables due to the hadronic interactions after the QGP freezes out will be deferred to the next section.

Because of the medium modification, heavy flavor hadrons produced in nucleus-nucleus collisions display different spectra from those produced in proton-proton collisions. The two most widely utilized quantities that characterize the medium effect are the nuclear modification factor $R_\mathrm{AA}$ and the elliptic flow coefficient $v_2$:
\begin{align}
& R_\mathrm{AA}(p_\mathrm{T})\equiv\frac{1}{N_\mathrm{coll}}\frac{{dN^\mathrm{AA}}/{dp_\mathrm{T}}}{{dN^\mathrm{pp}}/{dp_\mathrm{T}}}, \\
& v_2(p_\mathrm{T})\equiv\langle \cos(2\phi)\rangle=\left\langle\frac{p_x^2-p_y^2}{p_x^2+p_y^2}\right\rangle,
\end{align}
which describe the overall energy loss and the asymmetric $p_\mathrm{T}$ modification of the probe particles respectively.

\begin{figure}[tb]
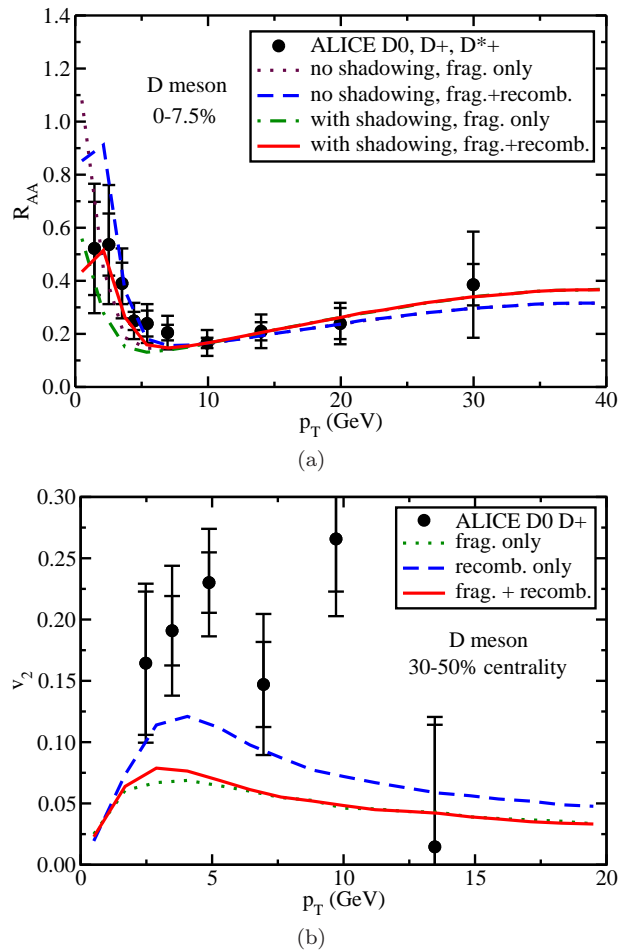

 \subfigure[]{\label{fig:RAA_DD5-LHC-0-7d5}
      \epsfig{file=plot-RAA_DD5-LHC-0-7d5.eps, width=0.45\textwidth, clip=}}
 \subfigure[]{\label{fig:v2_LHC-30-50}
      \epsfig{file=plot-v2_LHC-30-50.eps, width=0.45\textwidth, clip=}}
 \caption{(Color online) The $D$ meson (a) $R_\mathrm{AA}$ and (b) $v_2$ in 2.76~TeV Pb-Pb collisions \cite{Grelli:2012yv,Abelev:2013lca}, compared between different hadronization mechanisms, and between with and without the nuclear shadowing effect.}
 \label{fig:RAAv2LHC}
\end{figure}

In Fig. \ref{fig:RAA_DD5-LHC-0-7d5} we show our calculation of the $D$ meson $R_\mathrm{AA}$ in central Pb-Pb collisions at the LHC energy. The impact of the nuclear shadowing effect in heavy quark production in the initial state and the contribution of the coalescence mechanism to the $D$ meson formation can be clearly observed in the figure. As shown in Fig. \ref{fig:RAA_DD5-LHC-0-7d5}, if other factors are fixed, the inclusion of the initial state shadowing effect would lead to a factor of 2 suppression of the $D$ meson $R_\mathrm{AA}$ at low $p_\mathrm{T}$ and a mild enhancement at high $p_\mathrm{T}$. This is consistent with the findings shown in Fig. \ref{fig:LHCppvsAA}:  the production of charm quark is significantly suppressed at low $p_\mathrm{T}$ and slightly enhanced at high $p_\mathrm{T}$ in Pb-Pb collisions compared to that in proton-proton collisions. Therefore, a better understanding of the cold nuclear matter effect in the initial state is crucial for a more precise description of the heavy flavor suppression in nuclear collisions. From Fig. \ref{fig:RAA_DD5-LHC-0-7d5}, we also observe that although the fragmentation mechanism alone is sufficient for describing the heavy quark hadronization at high $p_\mathrm{T}$ (above 8~GeV), the coalescence of light and heavy quarks becomes crucial in the low and intermediate region: it converts low $p_\mathrm{T}$ heavy quarks into intermediate $p_\mathrm{T}$ hadrons by combining the former with thermal partons from the QGP medium, and thus suppresses the $D$ meson $R_\mathrm{AA}$ near zero $p_\mathrm{T}$ but greatly enhances it in between 2 and 5~GeV. With the incorporation of the nuclear shadowing effect in the initial state, a modified Langevin equation that includes both collisional and radiative energy loss of heavy quarks inside the QGP matter, and a hybrid model of fragmentation and coalescence, our calculation provides a good description of the $D$ meson $R_\mathrm{AA}$ in central Pb-Pb collisions as measured by the ALICE Collaboration. The spatial diffusion coefficient of heavy quark is determined as $5/(2\pi T)$ by comparing our calculation to experimental data at high $p_\mathrm{T}$, and will be utilized for all the following calculations in this section.

Figure \ref{fig:v2_LHC-30-50} shows our results of the $D$ meson $v_2$ in peripheral Pb-Pb collisions at the LHC. The nuclear shadowing effect is included for all the curves shown in this figure, but various hadronization mechanisms are compared in more details. For the pure fragmentation process, the Wigner function $f^W$ in the coalescence model is set as a constant 0 in order to switch off all coalescence channels; to the contrary, $f^W$ is fixed at 1 for the pure coalescence hadronization. One can observe that the pure coalescence limit leads to a much larger $D$ meson $v_2$ than the pure fragmentation limit because the former mechanism brings the anisotropic flow of light quarks from the hydrodynamic background into the formation of heavy mesons. However, only a slight enhancement in the $D$ meson $v_2$ at intermediate $p_\mathrm{T}$ is observed in our hybrid hadronization model compared to the pure fragmentation process despite the large enhancement of its yield. This may result from the momentum dependence of the Wigner function in this instantaneous coalescence model that prefers combining partons with similar velocities. Other factors may also affect the final $D$ meson $v_2$ such as the initial heavy quark spectra and the development of the radial flow in the hydrodynamic background.

This may result from a combinational effect of the initial heavy quark spectra, the momentum dependence of the Wigner function in this instantaneous coalescence model, and the development of the radial flow in the hydrodynamic background.

\begin{figure}[tb]
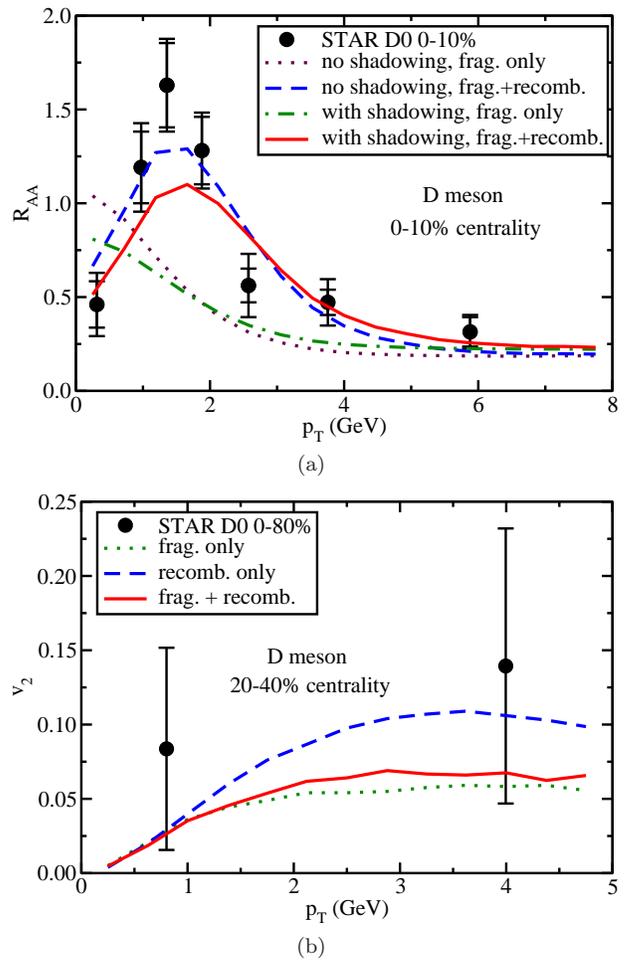

 \subfigure[]{\label{fig:RAA_DD5-RHIC-0-10}
      \epsfig{file=plot-RAA_DD5-RHIC-0-10.eps, width=0.45\textwidth, clip=}}
 \subfigure[]{\label{fig:v2_RHIC-20-40}
      \epsfig{file=plot-v2_RHIC-20-40.eps, width=0.45\textwidth, clip=}}
 \caption{(Color online) The $D$ meson (a) $R_\mathrm{AA}$ and (b) $v_2$ in 200~GeV Au-Au collisions \cite{Adamczyk:2014uip,Tlusty:2012ix}, compared between different hadronization mechanisms, and with and without the nuclear shadowing effect.}
 \label{fig:RAAv2RHIC}
 \end{figure}

In Fig. \ref{fig:RAAv2RHIC}, we study the suppression and the elliptic flow of $D$ mesons produced in the RHIC experiments. Although at the RHIC energy, the nuclear shadowing effect for the low $p_\mathrm{T}$ heavy quark is not as significant that at the LHC energy, it still has a non-negligible impact on the $D$ meson $R_\mathrm{AA}$ as shown in Fig. \ref{fig:RAA_DD5-RHIC-0-10}. Since the current RHIC experiments concentrate on the relatively low $p_\mathrm{T}$ region, the introduction of heavy-light quark coalescence is even more crucial in the hadronization process than that for describing the LHC data. The coalescence mechanism results in a bump structure of the $D$ meson $R_\mathrm{AA}$ around 1-2~GeV, which cannot be obtained with the pure fragmentation mechanism. In Fig. \ref{fig:v2_RHIC-20-40}, we can see that the introduction of the coalescence mechanism helps increase $D$ meson $v_2$, similar to the findings in the LHC scenario. By including all the effects discussed above, our numerical results are consistent with the STAR data.

\begin{figure}[tb]
  \epsfig{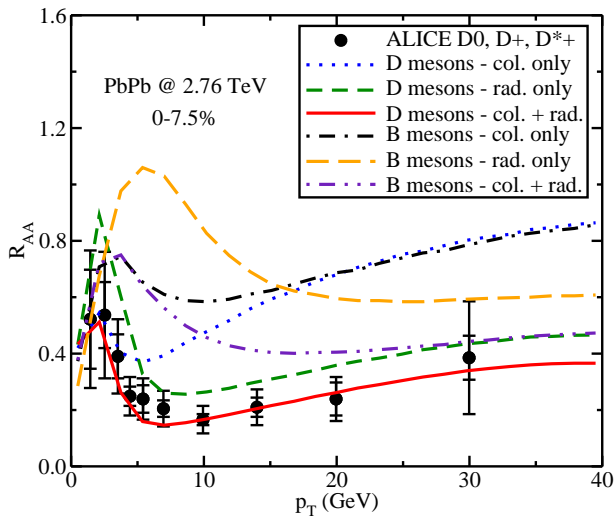}
  \caption{(Color online) Heavy meson suppression in central Pb-Pb collisions, compared between different energy loss mechanisms and $D$ and $B$ mesons.}
 \label{fig:plot-eloss-RAA-LHC-0-7d5}
\end{figure}

One of the most interesting puzzles related to heavy flavor is the mass hierarchy of parton energy loss. In Fig. \ref{fig:plot-eloss-RAA-LHC-0-7d5}, we compare the suppression between $D$ and $B$ mesons due to different energy loss mechanisms, in which the mass hierarchy of heavy quark energy loss can be clearly observed for both quasi-elastic scattering and medium-induced gluon radiation. Due to their larger masses, bottom quarks lose significantly smaller amount of energy than charm quark does at low $p_\mathrm{T}$ after they propagate through a realistic QGP medium and therefore $B$ meson displays larger $R_\mathrm{AA}$ than $D$ mesons. With our current model calculation, the mass effect on collisional energy loss becomes negligible for the meson spectra above 20~GeV. However, difference in radiative energy loss still remains up to 40~GeV. Apart from the mass hierarchy of the in-medium parton energy loss, there is also the mass dependence for heavy-light quark coalescence probability as shown in Fig. \ref{fig:Pcoal}. Since it is easier for bottom quarks to combine with thermal partons from the medium background than for charm quarks, the enhancement of the $B$ meson $R_\mathrm{AA}$ is more prominent than that of the $D$ meson $R_\mathrm{AA}$; such enhancement also spreads over a wider $p_\mathrm{T}$ regime for $B$ mesons.

\begin{figure}[tb]
  \epsfig{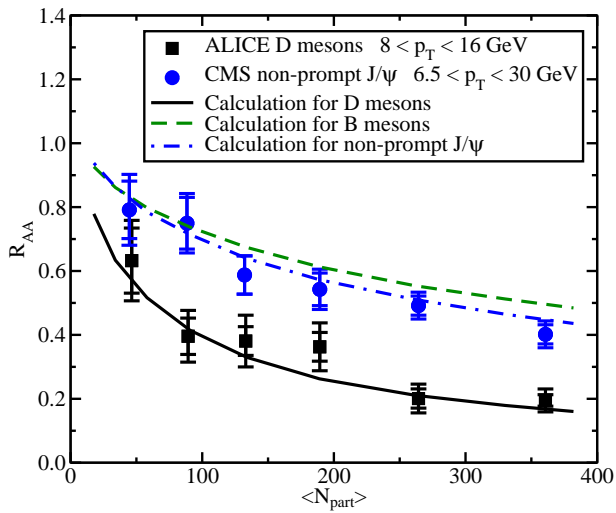}
  \caption{(Color online) Comparison of suppression between $D$, $B$ mesons and non-prompt $J/\psi$ in 2.76~TeV Pb-Pb collisions \cite{Bruna:2014pfa}.}
 \label{fig:RAA_LHC-npart}
\end{figure}

One possible direct verification of the mass hierarchy of parton energy loss is the comparison of the nuclear suppression of $D$ mesons versus non-prompt $J/\psi$ as shown in Fig. \ref{fig:RAA_LHC-npart}. Here we show our calculations of the participant number dependence of the $R_\mathrm{AA}$ for $D$ mesons, $B$ mesons and non-prompt $J/\psi$. The decay from $B$ meson to $J/\psi$ is implemented with \textsc{Pythia}. As has been mentioned earlier, the only free parameter in our transport model is the spatial diffusion coefficient of heavy quarks which is fixed to be $D=5/(2\pi T)$ by comparing high $p_\mathrm{T}$ $D$ meson $R_\mathrm{AA}$ in 2.76~TeV central Pb-Pb collisions to experimental data. One can see that with a single value for the transport coefficient, our calculation provides a good description of the participant number dependence of the suppression of $D$ meson and non-prompt $J/\psi$ simultaneously.

\section{Evolution of heavy mesons in a hadron gas}
\label{sec:hadronicInteraction}

As has been discussed in the previous section, at the critical temperature $T_\mathrm{c}$, both the QGP fireball and heavy quarks hadronize into color neutral bound states. We can obtain soft hadrons from the bulk matter via the Cooper-Frye formalism and obtain heavy hadrons through our hybrid hadronization model. Subsequently, the produced hadrons from each event are subject to hadronic rescattering which is modeled through UrQMD  \cite{Bass:1998ca,Bleicher:1999xi}.

Unlike the Langevin equation that only requires a single transport coefficient, the UrQMD model requires the microscopic cross sections of hadronic scatterings as crucial inputs. To simulate the rescatterings of $D$ mesons inside a hadron gas, we introduce into the UrQMD framework the scattering cross sections for charm mesons with $\pi$ and $\rho$ mesons as calculated in Refs. \cite{Lin:2000jp,Lin:1999ve,Lin:1999ad} which are based on a hadronic Lagrangian generated from local flavor SU(4) gauge symmetry. In this calculation, uncertainty remains in the choice of the cutoff parameter in the hadron form factors. We treat the variation in the cutoff as a systematic uncertainty in our following calculations of heavy meson observables.

 \begin{figure}[tb]
 \subfigure[]{\label{fig:HI-RAA_LHC-0-7d5}
      \epsfig{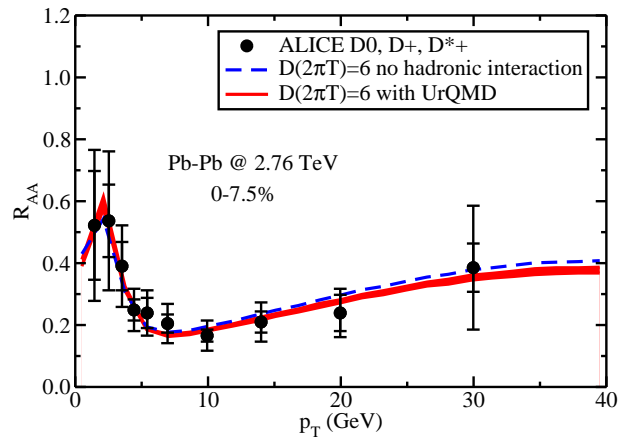}}
 \subfigure[]{\label{fig:HI-v2_LHC-30-50}
      \epsfig{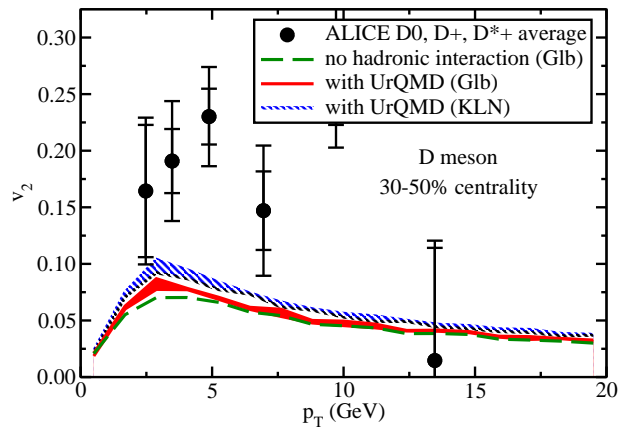}}
 \caption{(Color online) Effects of hadronic interactions on the $D$ meson (a) $R_\mathrm{AA}$ and (b) $v_2$ in 2.76~TeV Pb-Pb collisions.}
 \label{fig:RAAv2LHC-HI}
 \end{figure}
 
 \begin{figure}[tb]
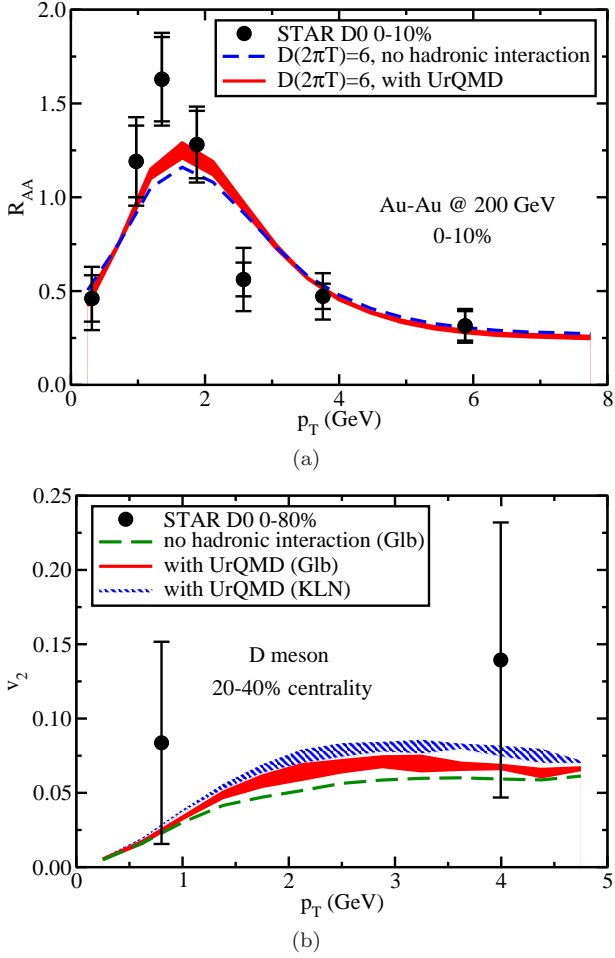

 \subfigure[]{\label{fig:HI-RAA_RHIC-0-10}
      \epsfig{file=plot-HI-RAA_RHIC-0-10.eps, width=0.45\textwidth, clip=}}
 \subfigure[]{\label{fig:HI-v2_RHIC-20-40}
      \epsfig{file=plot-HI-v2_RHIC-20-40.eps, width=0.45\textwidth, clip=}}
 \caption{(Color online) Effects of hadronic interactions on the $D$ meson (a) $R_\mathrm{AA}$ and (b) $v_2$ in 200~GeV Au-Au collisions.}
 \label{fig:RAAv2RHIC-HI}
 \end{figure}

In Fig. \ref{fig:HI-RAA_LHC-0-7d5}, we investigate how $D$ meson $R_\mathrm{AA}$ is affected by the hadronic interactions. One observes that due to the additional energy loss experienced by $D$ mesons inside the hadron gas, $R_\mathrm{AA}$ for $D$ mesons is further suppressed at large $p_\mathrm{T}$. Consequently, due to the conservation of the number of charmed hadrons, $D$ meson $R_\mathrm{AA}$ is slightly enhanced at low $p_\mathrm{T}$ after the UrQMD evolution. As mentioned above, the error bands in our results characterize the uncertainties introduced by a factor of 2 difference in the choice of the cutoff parameter in the hadron form factors when calculating the heavy meson scattering cross sections in Ref. \cite{Lin:2000jp}. With our comprehensive framework that incorporates heavy flavor evolution in both QGP and hadronic phases, we provide a good description of the $D$ meson suppression as observed in 2.76~TeV central Pb-Pb collisions. After the inclusion of the hadronic interactions, the spatial diffusion coefficient of heavy quarks extracted from high $p_\mathrm{T}$ $R_\mathrm{AA}$ data is updated to $6/(2\pi T)$.

The effect of the hadronic interactions on $D$ meson $v_2$ at the LHC energy is shown in Fig. \ref{fig:HI-v2_LHC-30-50}. Due to additional scatterings of $D$ mesons in an anisotropic hadron gas, its $v_2$ is further enhanced by around 20\%. In Fig. \ref{fig:HI-v2_LHC-30-50}, we also present the difference between two hydrodynamic initial conditions. Since the KLN model provides a larger eccentricity of the initial entropy density profiles than the Glauber model, this may cause another 20\% difference in the collective flow of heavy mesons after their evolutions inside the QGP and the hadron gas. However, after taking all effects into account, our calculation still underestimates $D$ meson $v_2$ compared to the ALICE data. Several studies has been carried out targeting this $v_2$ puzzle. For instance, it has been suggested in Refs. \cite{Xu:2014tda,Das:2015ana} that by taking into account the temperature dependence of the transport coefficient ($\hat{q}/T^3$) and increasing the relative contribution of the medium modification to heavy flavor spectra around $T_\mathrm{c}$, the anisotropy parameter $v_2$ in the final state can be effectively enhanced. These effects will be investigated in detail in the future.

In Fig. \ref{fig:RAAv2RHIC-HI}, we provide our calculations of the $D$ meson nuclear suppression factor and anisotropic flow parameter for Au-Au collisions at the RHIC energy. Similar to the LHC scenario, the hadronic interactions simulated with the UrQMD model slightly suppress $D$ meson $R_\mathrm{AA}$ at large $p_\mathrm{T}$ and enhance the anisotropy parameter $v_2$. Our numerical results are consistent with the experimental data from the STAR Collaboration.

\begin{figure}[tb]
  \epsfig{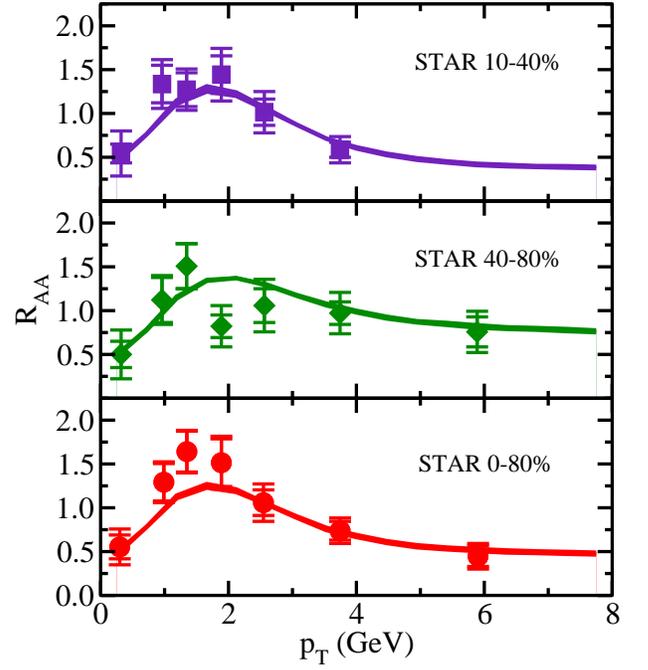}
  \caption{(Color online) The $D$ meson suppression in different centralities at the RHIC experiment.}
 \label{fig:HI-RAA_RHIC-cen}
\end{figure}

\begin{figure}[tb]
  \epsfig{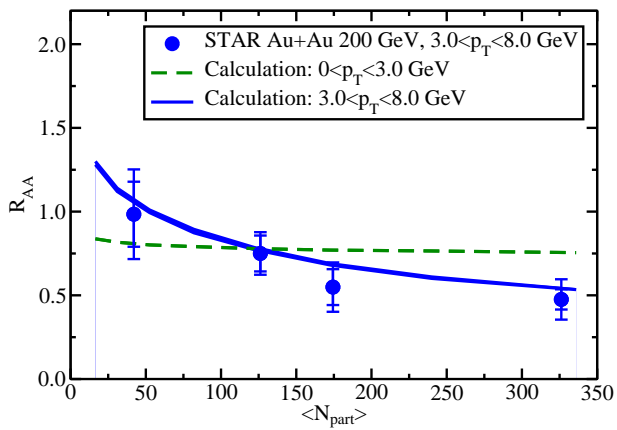}
  \caption{(Color online) The participant number dependence of the $D$ meson $R_\mathrm{AA}$ at the RHIC.}
 \label{fig:HI-RAA_RHIC-npart}
\end{figure}

In Fig. \ref{fig:HI-RAA_RHIC-cen}, we present $D$ meson $R_\mathrm{AA}$ for different centrality classes as measured in the RHIC experiments. In Fig. \ref{fig:HI-RAA_RHIC-npart}, we show the integrated $R_\mathrm{AA}$ of $D$ mesons over given $p_\mathrm{T}$ regions as a function of centrality characterized by the participant numbers. One can see that as moving from more central to more peripheral collisions, $D$ meson $R_\mathrm{AA}$ increases due to a smaller geometric size and a shorter lifetime of the hot and dense nuclear matter created in more peripheral collisions. Our calculations are consistent with all the available data from the RHIC and a prediction for the participant number dependence of the $D$ meson $R_\mathrm{AA}$ is also provided for a smaller $p_\mathrm{T}$ region. We note that the difference of the integrated $R_\mathrm{AA}$ between the $0<p_\mathrm{T}<3$~GeV regime and $3<p_\mathrm{T}<8$~GeV is expected to depend on the centrality of collisions due to a combination of heavy flavor energy loss and the coalescence mechanism in heavy meson production.

\section{Summary and outlook}
\label{sec:summary}

In this work, we have established a comprehensive framework to describe the full evolution history of heavy quarks together with the evolution of the fireball in relativistic heavy-ion collisions, including their initial production, energy loss in a QGP medium, hadronization an the subsequent interactions of heavy mesons with the hadron gas. 
At the beginning, the entropy density of the bulk matter produced in nuclear collisions is initialized with either the MC-Glauber or the MC-KLN model; heavy quarks are initialized with the MC-Glauber model for their position space distribution and a pQCD calculation for their momentum space distribution. 
During the QGP stage, the bulk matter evolves according to a (2+1)-dimensional viscous hydrodynamic model, while the heavy quark transport inside this medium is described by our modified Langevin equation incorporating both quasi-elastic scattering and medium-induced gluon radiation processes. 
At the critical temperature $T_\mathrm{c}$, the QGP matter is converted into soft hadrons according to the Cooper-Frye formalism, and heavy quarks on the other hand hadronize based on the hybrid model of fragmentation and coalescence model we develop. 
In the last stage, both soft and heavy hadrons are fed into the UrQMD model for the simulation of hadronic scatterings until all interactions cease. 
Our numerical framework is designed in such a way that each evolution stage can be easily replaced by another model, e.g., a different hydrodynamic background, a different heavy quark transport model or a different hadronization process, therefore a systematic comparison between different theoretic formalisms can be conveniently implemented in the future.

With our current approach, we have shown that while the collisional energy loss dominates the low $p_\mathrm{T}$ region of heavy quark transport inside the QGP, the contribution from the medium-induced gluon radiation is significant at high $p_\mathrm{T}$. During the hadronization process, the fragmentation mechanism dominates the high $p_\mathrm{T}$ regime, but the introduction of the heavy-light quark coalescence significantly enhances the production of heavy meson at intermediate $p_\mathrm{T}$ in nucleus-nucleus collisions. In addition, the hadronic interactions after the QGP decays further suppresses the heavy meson $R_\mathrm{AA}$ and enhances its elliptic flow $v_2$. 
In this work, the mass dependence of heavy flavor evolution is also investigated. It has been found that due to the larger mass of bottom quarks compared to charm quarks, the former lose less energy. 
The effect of such a mass hierarchy on the final heavy meson spectra fades away around $p_\mathrm{T}=20$~GeV for the collisional energy loss, but still remains up to 40~GeV for the radiative energy loss. 
Also due to the larger masses of bottom quarks, the coalescence dominates over a wider $p_\mathrm{T}$ region for the hadronization as compared to charm quarks. 
Within our framework, we have provided numerical results of the heavy meson suppression and anisotropic flow coefficients that are consistent with most data from both the LHC and the RHIC experiments. 
The spatial diffusion coefficient $D$ of heavy quark extracted from our model is between $5/(2\pi T)$ and $6/(2\pi T)$, depending on whether the hadronic interaction is included or not in the calculation. 
These numbers may be translated into $\hat{q}_\mathrm{A}/T^3$ around $9.4\sim11.3$ for a gluon jet, or $\hat{q}_\mathrm{F}/T^3$ around $4.2\sim5.0$ for a quark jet, which are consistent with the value obtained by the JET Collaboration by comparing various light flavor jet energy loss formalisms to the experimental data \cite{Burke:2013yra}.

Our study constitutes an important contribution towards a more quantitative and accurate understanding of the full evolution of heavy flavors produced in relativistic nuclear collisions. Several further improvements await our future effort. For instance, the nature of heavy quark dynamics in the pre-equilibrium stage of heavy-ion collisions \cite{Mrowczynski:1993qm,Das:2015aga} is still not clear at this moment, which may affect the final state hadron spectra. It has been suggested that heavy quarks produced by the gluon splitting process may experience different medium modification pattern compared to those directly produced through the hard scatterings \cite{Huang:2013vaa}. As discussed earlier, by increasing the relative contribution of energy loss near $T_\mathrm{c}$, the heavy quark $v_2$ may be increased a lot without affecting its overall suppression \cite{Xu:2014tda,Das:2015ana}; this might be helpful to explain the large $v_2$ puzzle. These aspects will be explored in our future work.

\section*{Acknowledgments}

We are grateful to the Ohio State University group (Z. Qiu, C. Shen, H. Song and U. Heinz) for providing the numerical codes of the hydrodynamical evolution and its initialization, and the Texas A\&M University group (K. C. Han, R. Fries, and C. M. Ko) for discussions on constructing the coalescence model. We also acknowledge the helpful advice from X.-N. Wang, and the computational resources provided by the Open Science Grid (OSG). This work is funded by the Director, Office of Energy Research, Office of High Energy and Nuclear Physics, Division of Nuclear Physics, of the U.S. Department of Energy under Contract Nos. DE-AC02-05CH11231 and DE-FG02-05ER41367, and within the framework of the JET Collaboration, and by the Natural Science Foundation of China (NSFC) under Grant No. 11375072.

\bibliographystyle{h-physrev5}
\bibliography{SCrefs}

\end{document}